\documentclass[a4paper]{llncs}
\usepackage[utf8]{inputenc}
\usepackage[T1]{fontenc}
\usepackage{imakeidx}
\usepackage{graphicx}
\usepackage{amssymb}
\usepackage{cwdefspie}
\usepackage{amssymb}
\usepackage{amsfonts}
\usepackage{amsmath}
\usepackage{xspace}
\usepackage{fancyvrb}
\usepackage{pdfpages}
\usepackage{wrapfig}
\usepackage[hidelinks,pdftitle=PIE]{hyperref}
\title{Facets of the \textit{PIE}
	    Environment\\ for Proving, Interpolating and Eliminating\\
	     on the Basis of First-Order Logic}
\author{Christoph Wernhard}
\institute{Berlin, Germany}
\newcommand{\assign}{\mathrel{\mathop:}=}

\newcommand{\f}[1]{\mathsf{#1}}
\newcommand{\true}{\top}
\newcommand{\false}{\bot}
\newcommand{\imp}{\rightarrow}
\newcommand{\revimp}{\leftarrow}
\newcommand{\equi}{\leftrightarrow}
\newcommand{\entails}{\models}

\newcommand{\pplmacro}[1]{\mathit{#1}}
\newcommand{\ppldefmacro}[1]{\mathit{#1}}
\newcommand{\pplparam}[1]{\mathit{#1}}

\newcommand{\pplparamplain}[1]{#1}
\newcommand{\pplparamplainidx}[2]{#1_{#2}}

\newcommand{\pplparamplainsup}[2]{#1^{#2}}

\makeatletter%
\newcommand{\entrymark}[1]{}%
\newcommand\entryhead{%
\@startsection{entry}{10}{\z@}{12pt plus 2pt minus 2pt}{0pt}{}}%
\makeatother
	    
\newcommand{\pplkbBefore}
{\entryhead*{}%
\setlength{\arraycolsep}{0pt}%
\pagebreak[0]%
\begin{samepage}%
\noindent%
\rule[0.5pt]{\textwidth}{2pt}\\%
\noindent}

\newcommand{\pplkbBetween}
{\setlength{\arraycolsep}{3pt}%
\\\rule[3pt]{\textwidth}{1pt}%
\par\nopagebreak\noindent Defined as\begin{center}}

\newcommand{\pplkbAfter}{\end{center}\end{samepage}\noindent}

\newcommand{\pplkbBodyBefore}{\par\noindent where\begin{center}}
\newcommand{\pplkbBodyAfter}{\end{center}}

\newcommand{\pplIsValid}[1]{\noindent This formula is valid: $#1$\par}

\makeindex

\begin{document}
  \maketitle

\makeatletter%
\renewcommand\entryhead{%
\@startsection{entry}{10}{\z@}{0pt plus 2pt}{0pt}{}}%
\makeatother
  
\newcommand{\PIE}{\textit{PIE}\xspace}
\newcommand{\CM}{\textit{CM}\xspace}
\newcommand{\code}[1]{{\small \texttt{#1}}}
\newcommand{\tvar}[1]{\textit{#1}}
\newsavebox{\fmbox}
\newenvironment{outbox}
               {\begin{center}
                 \noindent\begin{lrbox}{\fmbox}\begin{minipage}{0.95\textwidth}}
               {\end{minipage}\end{lrbox}\fbox{\usebox{\fmbox}}
                 \end{center}}
  
\begin{abstract} \PIE is a Prolog-embedded environment for automated reasoning
  on the basis of first-order logic.  Its main focus is on formulas, as
  constituents of complex formalizations that are structured through formula
  macros, and as outputs of reasoning tasks such as second-order quantifier
  elimination and Craig interpolation.  It supports a workflow based on
  documents that intersperse macro definitions, invocations of reasoners, and
  \LaTeX-formatted natural language text. Starting from various examples, the
  paper discusses features and application possibilities of \PIE along with
  current limitations and issues for future research.
  \end{abstract}
\section{Introduction}
\label{sec-intro}

First-order logic is used widely and in many roles in philosophy, mathematics,
and artificial intelligence as well as other branches of computer science.
Many practically successful reasoning approaches can be viewed as derived from
reasoning in first-order logic, for example, SAT solving, logic programming,
database query processing and reasoning in description logics.  The overall
aim of the \name{PIE} environment is to support the \emph{practical mechanized
reasoning in first-order logic}. Approaching this aim consequently leads from
first-order theorem proving in the strict sense to tasks that \emph{compute
first-order formulas}, in particular second-order quantifier elimination and
Craig interpolation, whose integrated support characterizes \PIE. The system
is written and embedded in \name{SWI-Prolog} \cite{swiprolog} and provides,
essentially as a library of Prolog predicates, a number of functionalities:

\begin{itemize}
  \item Support for a Prolog-readable syntax of first-order
logic formulas.  
\item Formula pretty-printing in Prolog syntax and in \LaTeX.
\item A versatile formula macro processor.  
\item Support for processing
documents that intersperse formula macro definitions, reasoner invocations and
\LaTeX-formatted natural language text.  
\item Interfaces to external first-order and propositional reasoners.  
\item A built-in Prolog-based first-order theorem prover.
\item Implemented reasoning techniques that compute formulas:
  \begin{itemize}
  \item Second-order quantifier elimination on the basis of first-order logic.
  \item Computation of first-order Craig interpolants.  
  \item Formula conversions for use in preprocessing, inprocessing and
output presentation.
  \end{itemize}
\end{itemize}  

\noindent
The system is available as free software from its homepage
 \begin{center} \url{http://cs.christophwernhard.com/pie}.  \end{center}
  The
  distribution includes several example documents whose source files as well
  as rendered \LaTeX\ presentations can also be accessed directly from the
  system Web page. \name{Inspecting Gödel’s
  Ontological Proof} is there an advanced application,
  where the interplay of elimination and modal axioms is
  applied in several contexts.  The system was first presented at the 2016
  workshop \name{Practical Aspects of Automated Reasoning} \cite{cw-paar}.
  Here we show various application possibilities, features and also issues for
  further research that become apparent with the system by starting from
  a number of examples. The paper is itself written as a \PIE document and thus
  includes fragments generated by \PIE and the included or integrated reasoners.
  
  The rest of this paper is structured as follows: After introducing in
  Sect.~\ref{sec-documents} the document-oriented workflow supported by \PIE,
  we show in Sect.~\ref{sec-soqe} how it applies to the
  invocation of second-order
  quantifier elimination in the system. Section~\ref{sec-soqe-abduction}
  provides an application example of elimination, a certain form of abduction,
  which is shown together with basic features of the \PIE macro system. We
  proceed in Sect.~\ref{sec-tp} to outline how systems for theorem proving in
  the strict sense are embedded into \PIE. In Sect.~\ref{sec-circ} the
  computation of circumscription is discussed as another example of
  second-order quantifier elimination with \PIE, along with further features
  of the macro system and the general issue of finding good presentations of
  computed formulas that are essentially just characterized semantically.
  Section~\ref{sec-colorability} sketches a further application of second-order
  quantifier elimination: a potential way of logic programming with
  second-order formulas as used for theoretical considerations in descriptive
  complexity. Further features of \PIE are summarized in
  Sect.~\ref{sec-further}, and Sect.~\ref{sec-conclusion} concludes the
  paper.

  Related work is discussed in the respective contexts. The bibliography is
  somewhat extensive, reflecting that the system relates to methods as well as
  implementation and application aspects in a number of areas, including
  first-order theorem proving, Craig interpolation, second-order quantifier
  elimination and knowledge representation.
\section{\PIE Documents}
\label{sec-documents}

The main way to interact with \PIE is by developing or modifying a \name{\PIE
  document}, a file that intersperses definitions of formula macros,
  specifications of reasoning tasks, and \LaTeX-formatted natural language
  text in the fashion of literate programming \cite{knuth:literate}.  Such a
  document can be \emph{loaded} into the Prolog environment like a source code
  file. Reasoner invocations, where the defined macros are available, can then
  be submitted as inputs on the Prolog console.  The document can also be
  \emph{processed}, which results in a generated \LaTeX\ document: Macro
  definitions are pretty-printed in \LaTeX, specified reasoner invocations are
  executed and a pretty-printed \LaTeX\ result presentation is inserted, and
  \LaTeX\ fragments are inserted directly.  The generated \LaTeX\ document can
  then be displayed in PDF format.

  Aside of indentation, the \LaTeX\ pretty-printer can apply certain symbol
  conversions to subscripted or primed symbols.  Also a compact syntax where
  parentheses to separate arguments from functors and commas between arguments
  are omitted is available as an option for both Prolog and \LaTeX\ forms.

  \PIE source documents can be re-loaded into the Prolog environment such that
  mechanized formalizations can be developed in a workflow similar to
  programming in AI languages like Prolog and Lisp.
  
  First-order reasoners are often heavily dependent on configuration
  settings. A \name{\PIE document} specifies all information needed to
  reproduce the results of reasoner invocations in a convenient way. Effective
  configuration parameters are combined from system defaults, defaults declared
  in the document and options supplied with particular specifications of
  reasoner invocations.

\section{Second-Order Quantifier Elimination in \PIE}
\label{sec-soqe}
  
Second-order quantifier elimination is the task of computing for a given
  formula with second-order quantifiers, that is, quantifiers upon predicate
  or function symbols, an equivalent first-order formula.  \PIE so far just
  supports second-order quantification upon predicate symbols, or
  \defname{predicate quantification}.  Here is an example
of \PIE's  \LaTeX\ representation of the invocation of a reasoner that
performs second-order quantifier elimination:
\par\medskip

\noindent Input: $\exists \mathit{p} \, (\forall \mathit{x} \, (\mathsf{q}(\mathit{x}) \imp  \mathit{p}(\mathit{x})) \land  \forall \mathit{x} \, (\mathit{p}(\mathit{x}) \imp  \mathsf{r}(\mathit{x}))).$\\
\noindent Result of elimination:
\[\begin{array}{lllll}
\forall \mathit{x} \, (\mathsf{q}(\mathit{x}) \imp  \mathsf{r}(\mathit{x})).
\end{array}
\]
The source code in the \PIE document that effects this output is:

\medskip

\noindent  
\begin{BVerbatim}[fontsize=\small]
:- ppl_printtime(ppl_elim(ex2(p, (all(x, (q(x) -> p(x))),
				  all(x, (p(x) -> r(x))))))).
\end{BVerbatim}

\medskip \noindent The directive \code{ppl\_printtime} effects that its
  argument is evaluated at ``print time'', that is, at \name{processing}, when
  the \LaTeX\ presentation is generated.\footnote{The prefix \code{ppl\_} of
  this and related predicates should suggest \name{pretty-print in \LaTeX\
  format.}}  The argument is an invocation of the elimination reasoner with
  the predicate $\code{ppl\_elim}$. It has a formula as argument, possibly
  with predicate quantifiers. If called at ``print time'' it prints inputs and
  outputs formatted in \LaTeX\, as shown above for the example. It can also be
  invoked in the context of plain Prolog processing, where it just effects
  that the output is pretty printed in Prolog syntax. The following
  interaction would, for example, be possible in the Prolog console:

  \medskip
  
\noindent  
\begin{BVerbatim}[fontsize=\small]  
?- ppl_elim(ex2(p, (all(x, (q(x) -> p(x))), all(x, (p(x) -> r(x)))))).
all(x, (q(x)->r(x)))
true.
\end{BVerbatim}

  \medskip

\noindent Printing the output is performed there as a side effect.
  \name{SWI-Prolog} afterwards prints $\code{true.}$ to indicate that the
  invocation of $\code{ppl\_elim}$ was successful. To access the output
  formula from a program, \PIE provides two alternate means: With an option
  list \code{[printing=false, r=Result]} as second argument,
  $\code{ppl\_elim}$ does not effect that the elimination result is print\-ed,
  but instead bound to the Prolog variable \code{Result} for further
  processing. The second way to access the result formula of the last reasoner
  invocation is with the supplied predicate
  \code{last\_ppl\_result(\textit{Result})}.  This predicate may itself be
  used in macro definitions.

  Let us take a brief look at the syntax of the argument formula of
  \code{ppl\_elim} in the example. It represents a second-order formula as a
  Prolog ground term.  Conjunction is represented as in Prolog by
  $\code{,}/2$ and implication by $\code{->}/2$, with standard
  operator settings from Prolog. The universal first-order quantifier is
  expressed by $\code{all}/2$ and the existential second-order quantifier by
  \code{ex2}/2.
  
  \PIE performs second-order quantifier elimination by an included Prolog
  implementation of the \name{DLS} algorithm \cite{dls}, a method based on
  formula rewriting until second-order subformulas have a certain shape that
  allows elimination in one step by rewriting with Ackermann's lemma, an
  equivalence due to \cite{ackermann:35}.  Implementing \name{DLS} brings
  about many subtle and interesting issues
  \cite{dls:gustafsson,dls:conradie,cw-relmon}, for example, incorporation of
  non-deterministic alternative courses, dealing with un-Skolemization,
  simplification of formulas in non-clausal form and ensuring success of the
  method for certain input classes.  The current implementation in \PIE is far
  from optimum solutions of these issues, but can nevertheless be used in
  nontrivial applications and might contribute to improvements by making
  experiments possible.
  
  Of course, second-order quantifier elimination on the basis of first-order
  logic does not succeed in general. Nevertheless, along with variants termed
  \name{forgetting}, \name{uniform interpolation} or \name{projection}, it has
  many applications, including deciding fragments of first-order logic
  \cite{loewenheim:15,beh:22}, computation of frame correspondence properties
  from modal axioms \cite{scan,dls,schmidt:2012:ackermann}, computation of
  circumscription \cite{dls}, embedding nonmonotonic semantics in a classical
  setting \cite{cw-projcirc,cw-logprog}, abduction with respect to classical
  and to nonmonotonic semantics \cite{lin:01:snc,dls-snc,cw-abduction},
  forgetting in knowledge bases
  \cite{lin-forget,cw-skp,ks:2013:frocos,ludwig:dl14,delgrande:17}, and
  approaches to modularization of knowledge bases derived from the notion of
  conservative extension
  \cite{dl-conservative,grau:modular:2008,lutz:ijcai11}.  Further applications
  of second-order quantifier elimination are described in the monograph
  \cite{soqe}.
  
  For second-order quantifier elimination and similar operations there are
  several implementations based on modal and description logics, but very few
  on first-order logic: A Web service\footnote{Available at
  \url{http://www.mettel-prover.org/scan/}.} invokes an implementation
  \cite{scan:engel} of the \name{SCAN} algorithm \cite{scan}.
  \name{DLSForgetter} \cite{alassaf:schmidt:19} is a recent system that
  implements the \name{DLS} algorithm \cite{dls}.  An earlier implementation
  \cite{dls:gustafsson} of \name{DLS} seems to be no longer available.
  
\section{Abduction with Second-Order Quantifier Elimination --
Basic Use of \PIE Macros}
\label{sec-soqe-abduction}

In the simplest case, a \PIE formula macro serves as a formula label that may
be used in subformula position in other formulas and is expanded into its
definiens. Here is an example of such a \PIE macro definition in the \LaTeX\
presentation:\par
\pplkbBefore
\index{kb1@$\ppldefmacro{kb_{1}}$}$\begin{array}{lllll}
\ppldefmacro{kb_{1}}
\end{array}
$\pplkbBetween
$\begin{array}{lllll}
(\mathsf{sprinkler\_was\_on} \imp  \mathsf{wet}(\mathsf{grass})) &&&&\; \land \\
(\mathsf{rained\_last\_night} \imp  \mathsf{wet}(\mathsf{grass})) &&&&\; \land \\
(\mathsf{wet}(\mathsf{grass}) \imp  \mathsf{wet}(\mathsf{shoes})).
\end{array}
$\pplkbAfter
\par\smallskip
\noindent The corresponding source is:
\par\medskip

\noindent
\begin{BVerbatim}[fontsize=\small]
def(kb1) ::
(sprinkler_was_on -> wet(grass)),
(rained_last_night -> wet(grass)),
(wet(grass) -> wet(shoes)).
\end{BVerbatim}

\medskip

\noindent The source statement has the form \code{def(\textit{MacroName}) ::
  \textit{ExpansionFormula}.}, where \code{::} is an infix operator with lower
  precedence than the operators used as connectives for logical formulas.
  Formula $\mathit{kb}_1$ is now defined as a small knowledge base that
  expresses a variant of a scenario often used to illustrate abduction.
  Actually, we use it now to show how a certain form of computing abductive
  explanations can be considered as second-order quantifier elimination.  It
  is based on the notion of \name{weakest sufficient condition}
  \cite{lin:01:snc,dls-snc,cw-projcirc}, which is basically a second-order
  formula that expresses the weakest formula in a given vocabulary that needs
  to be conjoined to given axioms to make a given theorem candidate an actual
  theorem.  This second-order formula as such is not very informative as it
  contains the axioms and the theorem as constituents, with disallowed symbols
  bound by quantifiers and possibly renamed but still present. However, the
  result of applying elimination to that second-order formula provides the
  weakest sufficient condition in the proper sense, or, considered with
  respect to abduction, the weakest explanation.

  \PIE allows to specify macros with parameters that are represented by Prolog
  variables. We utilize this to specify schematically the weakest explanation
  (or weakest sufficient condition) of observation $\begin{array}{l}
\pplparam{Obs}
\end{array}
$ on the
  complement of $\begin{array}{l}
\pplparam{Na}
\end{array}
$ as assumables ($\begin{array}{l}
\pplparam{Na}
\end{array}
$ should
  suggest \name{\textbf{n}on-\textbf{a}ssumables}) within knowledge base
  $\begin{array}{l}
\pplparam{Kb}
\end{array}
$:
\pplkbBefore
\index{explanation(Kb,Na,Obs)@$\ppldefmacro{explanation}(\pplparam{Kb},\pplparam{Na},\pplparam{Obs})$}$\begin{array}{lllll}
\ppldefmacro{explanation}(\pplparam{Kb},\pplparam{Na},\pplparam{Obs})
\end{array}
$\pplkbBetween
$\begin{array}{lllll}
\forall \pplparam{Na} \, (\pplparam{Kb} \imp  \pplparam{Obs}).
\end{array}
$\pplkbAfter
\noindent The corresponding source code is:\par\medskip

\noindent
\begin{BVerbatim}[fontsize=\small]
def(explanation(Kb, Na, Ob)) ::
all2(Na, (Kb -> Ob)).
\end{BVerbatim}

\par\medskip\noindent \code{all2}/2 represents the universal second-order
  quantifier in \PIE's input formula syntax. The first argument of \code{all2}
  specifies the quantified predicates, either as a single Prolog atom or as
  list of atoms. In the example, there is the macro parameter
  $\begin{array}{l}
\pplparam{Na}
\end{array}
$ that needs to be instantiated correspondingly when the
  macro is expanded.
  The expression
  $\begin{array}{l}
\pplmacro{explanation}(\pplmacro{kb_{1}},{[}\mathsf{wet}{]},\mathsf{wet}(\mathsf{shoes}))
\end{array}
$ expands into the
  following ``non-informative'' version of the weakest sufficient condition:
\[\begin{array}{lllll}
\forall \mathit{p} \, ((\mathsf{sprinkler\_was\_on} \imp  \mathit{p}(\mathsf{grass})) &&&\; \land \\
\hphantom{\forall \mathit{p} \, (} (\mathsf{rained\_last\_night} \imp  \mathit{p}(\mathsf{grass})) &&&\; \land \\
\hphantom{\forall \mathit{p} \, (} (\mathit{p}(\mathsf{grass}) \imp  \mathit{p}(\mathsf{shoes})) &&&&\; \imp \\
\hphantom{\forall \mathit{p} \, (} \mathit{p}(\mathsf{shoes})).
\end{array}
\]
  
\noindent Second-order quantifier elimination applied to this formula yields
  the proper weakest explanation for the observation \code{wet(shoes)} in
  which the predicate $\code{wet}$ itself does not occur, with respect to the
  background knowledge base $\mathit{kb}_1$:

  \medskip

\noindent Input: $\pplmacro{explanation}(\pplmacro{kb_{1}},{[}\mathsf{wet}{]},\mathsf{wet}(\mathsf{shoes})).$\\
\noindent Result of elimination:
\[\begin{array}{lllll}
\mathsf{rained\_last\_night} \lor  \mathsf{sprinkler\_was\_on}.
\end{array}
\]
 \noindent
  It was obtained by the following directive in the source
  document:

  \medskip
  
\noindent
\begin{BVerbatim}[fontsize=\small]
:- ppl_printtime(ppl_elim(explanation(kb1,[wet],wet(shoes)))).
\end{BVerbatim}

\medskip
  
\noindent In \cite{cw-abduction} this approach to abduction has been
  generalized to non-monotonic semantics of logic programming, including the
  three-valued partial stable models semantics.
  \section{Invoking Theorem Provers from \PIE}
  \label{sec-tp}
  
  The abductive explanation computed in the previous section can be validated
  with a theorem prover. The presentation of the prover invocation and the
  result is in \PIE as follows:\par\medskip
\pplIsValid{\pplmacro{kb_{1}} \land  (\mathsf{rained\_last\_night} \lor  \mathsf{sprinkler\_was\_on}) \imp  \mathsf{wet}(\mathsf{shoes}).}
\par\medskip\noindent The corresponding source directive is

\medskip\noindent  
\begin{BVerbatim}[fontsize=\small]  
:- ppl_printtime(ppl_valid((kb1, (rained_last_night ; sprinkler_was_on)
			   -> wet(shoes)))).
\end{BVerbatim}
\par\medskip\noindent  
The semicolon \code{;}/2 represents disjunction, as in Prolog.  The reasoner
invocation predicate $\code{ppl\_valid}$ by default first calls the model
searcher \name{Mace4} with a short timeout, and, if it can not find a
``counter''-model of the negated formula, calls the prover \name{Prover9},
again with a short timeout.\footnote{\name{Prover9} and \name{Mace4} were
developed between 2005 and 2010 by William McCune. Their homepage is
\url{https://www.cs.unm.edu/~mccune/prover9/}.} Correspondingly,
$\code{ppl\_valid}$ prints a representation of one of three result values:
\name{valid}, \name{not valid} or \name{failed to validate} and in \LaTeX\
``print time'' mode also the input formula, as shown above.

  Like $\code{ppl\_elim}$, also $\code{ppl\_valid}$ can be called with a list
  of options as second argument. This allows to obtain Prolog term
  representations of \name{Prover9}'s resolution proof or \name{Mace4}'s
  model, to skip the call to \name{Mace4}, modify the configuration of
  \name{Mace4} and \name{Prover9}, or to specify another theorem prover to be
  called.

  Other provers can be incorporated through a generic interface to the
  \name{TPTP} \cite{tptp} syntax for proving problems, supported by most
  current first-order provers. In addition, \name{DIMACS} and \name{QDIMACS},
  the common formats of SAT and QBF solvers, respectively, are supported by
  \PIE.  Large propositional formulas are handled there efficiently with an
  internal representation implemented with destructive term operations.  Most
  of the support of propositional formulas is inherited from the precursor
  system \name{ToyElim} \cite{cw-toyelim}.

  \PIE also includes a Prolog-based first-order prover, \name{CM,} whose
  calculus can be understood as model elimination, clausal tableau
  construction \cite{handbook:tableaux:letz}, or the connection method
  \cite{bibel:1983}, similar to provers of the \name{leanCoP} family
  \cite{leancop,kaliszyk15:tableaux,femalecop}.  Its implementation follows
  the compilation-based \name{Prolog Technology Theorem Prover (PTTP)}
  paradigm \cite{pttp}. It computes proofs that are represented by Prolog
  terms and can be used to compute Craig interpolants
  (Sect.~\ref{sec-craig}). Details and evaluation results are available at
  \url{http://cs.christophwernhard.com/pie/cmprover}.

\section{Computing Circumscription as Second-Order Quantifier Elimination --
  \PIE Macros with Prolog Bodies, Result Simplifications}
  \label{sec-circ}
  
  The circumscription of a predicate $P$ in a formula $F$ is a formula whose
  models are the models $I$ of $F$ that are minimal with respect to $P$. That
  is, there is no model $I'$ of $F$ that is like $I$ except that the extension
  of~$P$ in~$I'$ is a strict subset of the extension of~$P$ in~$I$.  Predicate
  circumscription can be expressed by a second-order schema such that the
  \emph{computation} of circumscription is second-order quantifier elimination
  \cite{dls}.  The second-order circumscription of predicate~$P$ in
  formula~$F$ can thus be defined as a \PIE macro as follows:
\pplkbBefore
\index{circ(P,F)@$\ppldefmacro{circ}(\pplparamplain{P},\pplparamplain{F})$}$\begin{array}{lllll}
\ppldefmacro{circ}(\pplparamplain{P},\pplparamplain{F})
\end{array}
$\pplkbBetween
$\begin{array}{lllll}
\pplparamplain{F} \land  \lnot  \exists \pplparamplainsup{P}{\prime} \, (\pplparamplainsup{F}{\prime} \land  \pplparamplainidx{T}{1} \land  \lnot  \pplparamplainidx{T}{2}),
\end{array}
$\pplkbAfter
\pplkbBodyBefore
$
\begin{array}{l}\pplparamplainsup{F}{\prime} \assign \pplparamplain{F}[\pplparamplain{P} \mapsto \pplparamplainsup{P}{\prime}],\\
\pplparamplain{A} \assign \mathrm{arity\ of }\; \pplparamplain{P}\; \mathrm{ in }\; \pplparamplain{F},\\
\pplparamplainidx{T}{1} \assign \mathrm{transfer\ clauses}\; {[}\pplparamplain{P}/\pplparamplain{A}\textrm{-}\mathsf{n}{]} \rightarrow {[}\pplparamplainsup{P}{\prime}{]},\\
\pplparamplainidx{T}{2} \assign \mathrm{transfer\ clauses}\; {[}\pplparamplainsup{P}{\prime}{]} \rightarrow {[}\pplparamplain{P}/\pplparamplain{A}\textrm{-}\mathsf{n}{]}.

\end{array}$\pplkbBodyAfter
  This definition utilizes that \PIE macro definitions may contain a Prolog
  body that permits expansions involving arbitrary computations. Utility
  predicates with pretty-printing templates for use in these bodies are
  provided for common tasks. The source of the above definition reads:
  \par\medskip
\noindent  
\begin{BVerbatim}[fontsize=\small]
def(circ(P, F)) ::
F, ~ex2(P_p, (F_p, T1, ~T2)) ::-
	mac_rename_free_predicate(F, P, pn, F_p, P_p),
	mac_get_arity(P, F, A),
	mac_transfer_clauses([P/A-n], p, [P_p], T1),
	mac_transfer_clauses([P/A-n], n, [P_p], T2).
\end{BVerbatim}
\medskip
  
\noindent The Prolog body is introduced with the \code{::-} operator, which is
  defined with a precedence between \code{::} and the operators used to
  represent logical formulas.  The unary operator \code{\~} represents
  negation in formulas.\footnote{The standard Prolog negation operator
  \code{\textbackslash +} is not suited to represent classical negation as it
  symbolizes $\not\vdash$, non-provability.}  The suffix \code{\_p} used for
  some variable names is translated to the prime superscript in the \LaTeX\,
  rendering.  We only indicate here the effects of the auxiliary predicates in
  the Prolog body with an example: The formula $\begin{array}{l}
\pplmacro{circ}(\mathsf{p},\mathsf{p}(\mathsf{a}))
\end{array}
$
  expands into:
\[\begin{array}{lllll}
\mathsf{p}(\mathsf{a}) &&&&\; \land \\
\lnot  \exists \mathit{q} \, (\mathit{q}(\mathsf{a}) \land  \forall \mathit{x} \, (\mathit{q}(\mathit{x}) \imp  \mathsf{p}(\mathit{x})) \land  \lnot  \forall \mathit{x} \, (\mathsf{p}(\mathit{x}) \imp  \mathit{q}(\mathit{x}))).
\end{array}
\]
 \par\noindent Second-order quantifier elimination can be applied
    to compute the circumscription for the example:\par\medskip

\noindent Input: $\pplmacro{circ}(\mathsf{p},\mathsf{p}(\mathsf{a})).$\\
\noindent Result of elimination:
\[\begin{array}{lllll}
\mathsf{p}(\mathsf{a}) \land  \forall \mathit{x} \, (\mathsf{p}(\mathit{x}) \imp  \mathit{x}=\mathsf{a}).
\end{array}
\]
 \par\medskip\noindent As a more complex example, we consider
  circumscribing $\begin{array}{l}
\mathsf{wet}
\end{array}
$ in $\begin{array}{l}
\pplmacro{kb_{1}}
\end{array}
$:\par\medskip

\noindent Input: $\pplmacro{circ}(\mathsf{wet},\pplmacro{kb_{1}}).$\\
\noindent Result of elimination:
\[\begin{array}{lllll}
(\mathsf{rained\_last\_night} \imp  \mathsf{wet}(\mathsf{grass})) &&&&\; \land \\
(\mathsf{sprinkler\_was\_on} \imp  \mathsf{wet}(\mathsf{grass})) &&&&\; \land \\
(\mathsf{wet}(\mathsf{grass}) \imp  \mathsf{wet}(\mathsf{shoes})) &&&&\; \land \\
\forall \mathit{x} \, (\mathsf{wet}(\mathit{x}) \imp  \mathsf{rained\_last\_night} \lor  \mathsf{sprinkler\_was\_on}) &&&&\; \land \\
\forall \mathit{x} \, (\mathsf{wet}(\mathit{x}) \land  \mathsf{wet}(\mathsf{grass}) \imp  \mathit{x}=\mathsf{grass} \lor  \mathit{x}=\mathsf{shoes}).
\end{array}
\]
  \noindent The first three implications of this output form the expansion
  of~$\begin{array}{l}
\pplmacro{kb_{1}}
\end{array}
$.  The last two implications are added by the
  circumscription.  This particular form was actually obtained by applying a
  certain simplification to the formula returned directly by the elimination
  method:

  \medskip
\noindent  
\begin{BVerbatim}[fontsize=\small]
:- ppl_printtime(ppl_elim(circ(wet,kb1), [simp_result=[c6]])).
\end{BVerbatim}

  \medskip \noindent The option \code{[simp\_result=[c6]]} supplied to
  \code{ppl\_elim} effects that the elimination result is postprocessed by
  equivalence preserving conversions with the aim to make it more readable.
  The conversion named \code{c6} chosen for this example converts to
  conjunctive normal form, applies various clausal simplifications and then
  converts back to a quantified first-order formula, involving
  un-Skolemization if required. That the last conjunct of the result can be
  replaced by the more succinct $\forall \mathit{x} \,
  (\mathsf{wet}(\mathit{x}) \imp \mathit{x}=\mathsf{grass} \lor
  \mathit{x}=\mathsf{shoes})$ is, however, not detected by the current
  implementation.

  Finding good presentations of formulas, in particular in presence of
  operations that yield formulas with essentially semantic characterizations,
  is a challenging topic in general.

\section{Expressing Graph Colorability by a Second-Order Formula --
  \PIE Macros with Parameters in Functor Position}
  \label{sec-colorability}
  One of the fundamental results of descriptive complexity is the equivalence
  of NP and expressibility by an existential second-order formula
  (with respect to finite models), that is, a first order formula prefixed
  with existential predicate quantifiers. This view allows, for example, to
  specify 2-colorability\footnote{3-colorability, which is NP-complete, can be
  specified analogously. We consider here 2-colorability for brevity of the
  involved formulas.}  with respect to a relation~$E$ that specifies a graph
  as follows:\par
\pplkbBefore
\index{col2(E)@$\ppldefmacro{col_{2}}(\pplparamplain{E})$}$\begin{array}{lllll}
\ppldefmacro{col_{2}}(\pplparamplain{E})
\end{array}
$\pplkbBetween
$\begin{array}{lllll}
\exists \mathit{r} \exists \mathit{g} \, (\forall \mathit{x} \, (\mathit{r}(\mathit{x}) \lor  \mathit{g}(\mathit{x})) &&&&\; \land \\
\hphantom{\exists \mathit{r} \exists \mathit{g} \, (} \forall \mathit{x} \forall \mathit{y} \, (\pplparamplain{E}(\mathit{x},\mathit{y}) \imp  \lnot  (\mathit{r}(\mathit{x}) \land  \mathit{r}(\mathit{y})) \land  \lnot  (\mathit{g}(\mathit{x}) \land  \mathit{g}(\mathit{y})))).
\end{array}
$\pplkbAfter
\smallskip\noindent  
The source of this definition is:\par\medskip
\noindent  
\begin{BVerbatim}[fontsize=\small]  
def(col2(E)) ::
ex2([r,g],
    ( all(x, (r(x) ; g(x))),
      all([x,y], (E(x,y) -> (~((r(x), r(y))), ~((g(x), g(y)))))))).
\end{BVerbatim}
\medskip

\noindent The macro parameter \code{E} appears as a Prolog variable in
  predicate position.\footnote{\name{SWI-Prolog} can be configured to permit
  variable names as functors, which are read in as atoms with capitalized
  names.  The macro processor of \PIE compares them to actual variable names
  in the macro definition.}  The macro can then be used with instantiating
  \code{E} to a predicate symbol, or to a $\lambda$-expression that describes a
  particular graph (we will see examples in a moment).

  Specifying algorithms as (existential) second-order formulas seems very
  elegant, but so far not established as a \emph{practical} approach to logic
  programming. \PIE in its current implementation lets become apparent related
  desiderata: Instantiation with a predicate symbol should be usable as basis
  for abstract reasoning.  Instantiation with a $\lambda$-expression (or
  conjoining a definition of a graph), should permit successful elimination.
  If adequate, the problem should then automatically be converted to a form
  that can be processed by a SAT solver.

  So far, in the current implementation of \PIE, such steps just work in part,
  e.g., by decomposing the overall task manually into intermediate steps with
  different manually controlled formula simplifications, as illustrated by
  the following example. The following macro defines the inner, first-order,
  component of the above specification of 2-colorability:\par
\pplkbBefore
\index{fo_col2(E)@$\ppldefmacro{fo\_col_{2}}(\pplparamplain{E})$}$\begin{array}{lllll}
\ppldefmacro{fo\_col_{2}}(\pplparamplain{E})
\end{array}
$\pplkbBetween
$\begin{array}{lllll}
\forall \mathit{x} \, (\mathsf{r}(\mathit{x}) \lor  \mathsf{g}(\mathit{x})) &&&&\; \land \\
\forall \mathit{x} \forall \mathit{y} \, (\pplparamplain{E}(\mathit{x},\mathit{y}) \imp  \lnot  (\mathsf{r}(\mathit{x}) \land  \mathsf{r}(\mathit{y})) \land  \lnot  (\mathsf{g}(\mathit{x}) \land  \mathsf{g}(\mathit{y}))).
\end{array}
$\pplkbAfter
 \noindent \PIE allows to instantiate~$E$ in $\mathit{fo\_col2}(E)$ with a
  predicate constant~$\f{e}$ and eliminate one of the color
  predicates:\footnote{One color predicate can also be eliminated from an
  analogous specification of 3-co\-lorability.}
  \par\medskip

\noindent Input: $\exists \mathit{g} \, \pplmacro{fo\_col_{2}}(\mathsf{e}).$\\
\noindent Result of elimination:
\[\begin{array}{lllll}
\forall \mathit{x} \forall \mathit{y} \, (\mathsf{e}(\mathit{x},\mathit{y}) \imp  \lnot  (\mathsf{r}(\mathit{y}) \land  \mathsf{r}(\mathit{x})) \land  (\mathsf{r}(\mathit{y}) \lor  \mathsf{r}(\mathit{x}))).
\end{array}
\]
 \par\medskip\noindent 2-colorability for a given graph represented by a
  $\lambda$-expression can be evaluated by \PIE currently just in two steps
  with different elimination configurations, as performed by the
  following Prolog predicate:\par\medskip

\noindent  
\begin{BVerbatim}[fontsize=\small]  
elim_col2(E) :-
	ppl_elim(ex2([g], fo_col2(E)),
		 [elim_options=[pre=[c6]], printing=false, r=F1]),
	ppl_elim(ex2([r], F1),
		 [elim_options=[pre=[d6]], printing=false, r=F2]),
        ppl_form(E),
	ppl_form(F2).
\end{BVerbatim}
\par\medskip
\noindent
Options \code{printing=false} suppress the emission of printed representations
of the two invocations of the elimination reasoner. Only the
input $\lambda$-expression and the
final result are pretty-printed with calls to \code{ppl\_form}.
Options \code{pre=[c6]} and
\code{pre=[d6]} effect that preprocessing based on conversion to CNF and DNF,
respectively, is applied for elimination. Invoking\par\medskip

\noindent  
\begin{BVerbatim}[fontsize=\small]    
:- ppl_printtime(elim_col2(lambda([u,v],((u=1,v=2); (u=2,v=3))))).  
\end{BVerbatim}
\par\medskip
\noindent  
yields the following output:
\par
\[\begin{array}{lllll}
\lambda (\mathit{u},\mathit{v}).(\mathit{u}=\mathsf{1} \land  \mathit{v}=\mathsf{2}) \lor  (\mathit{u}=\mathsf{2} \land  \mathit{v}=\mathsf{3}).
\end{array}
\]
\[\begin{array}{lllll}
\mathsf{1}\neq \mathsf{2} \land  \mathsf{2}\neq \mathsf{3}.
\end{array}
\]
\par\medskip\noindent It expresses that the graph described by the
$\lambda$-expression is 2-colorable if and only if node~$1$ is not the same as
node~$2$ and node~$2$ is not the same as node~$3$.

\section{Craig Interpolation}
\label{sec-craig}

By Craig's interpolation theorem \cite{craig:linear,craig:uses}, for given
first-order formulas $F$ and $G$ such that $F$ entails $G$ (or, equivalently,
$F \imp G$ is valid) a first-order formula~$H$ can be constructed such that
$F$ entails $H$, $H$ entails $G$ and $H$ contains only symbols (predicates,
functions, constants, free variables) that occur in both $F$ and $G$. \PIE
supports the computation of Craig interpolants~$H$, for given valid
implications $F \imp G$. Here is a propositional example:
\par\medskip

\noindent Input: $\mathsf{p} \land  \mathsf{q} \imp  \mathsf{p} \lor  \mathsf{r}.$\\
\noindent Result of interpolation:
\[\begin{array}{lllll}
\mathsf{p}.
\end{array}
\]

  \noindent  
The corresponding directive in the source document is:
\par\medskip  
\noindent  
\begin{BVerbatim}[fontsize=\small]  
:- ppl_printtime(ppl_ipol((p, q -> (p ; r)))).  
\end{BVerbatim}
\par\medskip

\noindent The predicate \code{ppl\_ipol} invokes the interpolation
  reasoner. It takes an implication $F \imp G$ as argument and,
  analogously to \code{ppl\_elim} (Sect.~\ref{sec-soqe}), prints an
  interpolant of $F$ and $G$.\footnote{In certain configurations it
  can also print several different interpolants.}
  Here is another example of Craig
  interpolation, where universal and existential
  quantification need to be combined:\footnote{\label{foot-craig-fol}This is
  an example which involves an inference step with a constant that
  occurs only on the left side ($\f{a}$) and a
  constant that occurs only on the right side ($\f{b}$), which
  can not be
  handled by certain
  resolution-based interpolation systems. See \cite{bonacina:15:on,kovacs:17}.
  In this particular example, the order of the quantifications in the result
  is not relevant.}
  \par\medskip

\noindent Input: $\forall \mathit{x} \, \mathsf{p}(\mathsf{a},\mathit{x}) \land  \mathsf{q} \imp  \exists \mathit{x} \, \mathsf{p}(\mathit{x},\mathsf{b}) \lor  \mathsf{r}.$\\
\noindent Result of interpolation:
\[\begin{array}{lllll}
\exists \mathit{x} \, \forall \mathit{y} \, \mathsf{p}(\mathit{x},\mathit{y}).
\end{array}
\]
  \noindent Craig interpolation has many applications in logics and
  philosophy, as already shown in \cite{craig:uses}.  Main applications in
  computer science are in verification \cite{mcmillan:handbook} and query
  reformulation, based on its relationship to definability and construction of
  definientia in terms of a given vocabulary
  \cite{toman:wedell:book,benedikt:book,benedikt:2017}. For these
  applications, actually interpolants that are further constrained, in
  dependency of further restrictions on the input formulas, are relevant.  We
  do not consider these here, but show how basic definability via Craig
  interpolation can be expressed in \PIE.

  A formula $G$ is called \defname{definable} in a formula $F$ \defname{in
  terms of} a set of predicates~$S$ if and only if there exists a formula $H$
  whose predicates are all in $S$ such that $F \entails G \equi H$.  The
  formula $H$ is then called a \defname{definiens} of $G$.
  Consider, for example, the following formula:
\pplkbBefore
\index{kb2@$\ppldefmacro{kb_{2}}$}$\begin{array}{lllll}
\ppldefmacro{kb_{2}}
\end{array}
$\pplkbBetween
$\begin{array}{lllll}
\forall \mathit{x} \, (\mathsf{p}(\mathit{x}) \imp  \mathsf{q}(\mathit{x}) \land  \mathsf{s}(\mathit{x})) &&&&\; \land \\
\forall \mathit{x} \, (\mathsf{s}(\mathit{x}) \imp  \mathsf{r}(\mathit{x})) &&&&\; \land \\
\forall \mathit{x} \, (\mathsf{q}(\mathit{x}) \land  \mathsf{r}(\mathit{x}) \imp  \mathsf{p}(\mathit{x})).
\end{array}
$\pplkbAfter
We can invoke a first-order prover from \PIE to verify that the formula
  $\f{p}(\f{a})$ is definable in $\mathit{kb}_2$ in terms of
  $\{\f{q},\f{r}\}$:\par\medskip
\pplIsValid{\pplmacro{kb_{2}} \imp  (\mathsf{p}(\mathsf{a}) \equi  \mathsf{q}(\mathsf{a}) \land  \mathsf{r}(\mathsf{a})).}
  \medskip\noindent
  Actually, since $\f{a}$ does not occur in $\mathit{kb}_2$, we
  can equivalently verify the following implication, whose
  right side is a universally quantified first-order definition:
  \par\medskip
\pplIsValid{\pplmacro{kb_{2}} \imp  \forall \mathit{a} \, (\mathsf{p}(\mathit{a}) \equi  \mathsf{q}(\mathit{a}) \land  \mathsf{r}(\mathit{a})).}
  \par\medskip\noindent
  We can now utilize the features of \PIE to formally characterize
  definability and synthesize definientia:
\pplkbBefore
\index{definiens(G,F,P)@$\ppldefmacro{definiens}(\pplparamplain{G},\pplparamplain{F},\pplparamplain{P})$}$\begin{array}{lllll}
\ppldefmacro{definiens}(\pplparamplain{G},\pplparamplain{F},\pplparamplain{P})
\end{array}
$\pplkbBetween
$\begin{array}{lllll}
\exists \pplparamplain{P} \, (\pplparamplain{F} \land  \pplparamplain{G}) \imp  \forall \pplparamplain{P} \, (\pplparamplain{F} \imp  \pplparamplain{G}).
\end{array}
$\pplkbAfter
  The interpolants of the left and right side of $\mathit{definiens}(G,F,P)$
  are exactly the definientia of $G$ in $F$ in terms of all predicates not in
  $P$.  The implication is valid if and only if definability holds.  The
  second-order quantifications in the implication are existential on the left
  and universal on the right side.\footnote{We actually encountered right side
  of the implication before in Sect.~\ref{sec-soqe-abduction} as the weakest
  sufficient condition in the macro definition of $\mathit{explanation}$.}
  Considering that an implication can be understood as disjunction of the
  \emph{negated} left side and the right side, if $F$ and $G$ are first-order,
  then $\mathit{definiens}(G,F,P)$ is a formula whose second-order quantifiers
  are all \emph{universal}. Such a second-order formula is valid if and only
  if the first-order formula obtained by renaming the quantified predicates
  with fresh symbols and dropping the second-order quantifiers is valid.  This
  translation is handled automatically by \PIE such that we can now we verify
  definability of $\f{p}(\f{a})$ by invoking a first-order prover from
  \PIE:\par\medskip
\pplIsValid{\pplmacro{definiens}(\mathsf{p}(\mathsf{a}),\pplmacro{kb_{2}},{[}\mathsf{p},\mathsf{s}{]}).}
 \par\medskip\noindent And, we can apply Craig interpolation to
  compute a definiens:\par\medskip

\noindent Input: $\pplmacro{definiens}(\mathsf{p}(\mathsf{a}),\pplmacro{kb_{2}},{[}\mathsf{p},\mathsf{s}{]}).$\\
\noindent Result of interpolation:
\[\begin{array}{lllll}
\mathsf{q}(\mathsf{a}) \land  \mathsf{r}(\mathsf{a}).
\end{array}
\]
  The implementation of the computation of Craig interpolants in \PIE operates
  by a novel adaption of Smullyan's interpolation method
  \cite{smullyan:book,fitting:book} to clausal tableaux \cite{cw-ipol}.
  Suitable clausal tableaux can be constructed by the Prolog-based prover \CM
  that is included in \PIE.  The system also supports the conversion of proof
  terms returned by the hypertableau prover \name{Hyper} \cite{cw-ekrhyper} to
  such tableaux and thus to interpolants, but this is currently at an
  experimental stage.\footnote{Hypertableaux, either obtained from a
  hypertableau prover or obtained from a clausal tableau prover like \CM by
  restructuring the tableau seem interesting as basis for interpolant
  extraction in query reformulation, as they allow to ensure that the
  interpolants are range restricted.  Some related preliminary results are in
  \cite{cw-ipol}.}

  The interpolants~$H$ constructed by \PIE strengthen the requirements for
  Craig interpolants in that they are actually Craig-Lyndon
  interpolants, that is,
  predicates occur in~$H$ only in polarities in which they occur in both~$F$
  and~$G$.  Symmetric interpolation \cite[Sect.~5]{mcmillan:symmetric} is
  supported in \PIE, implemented by computing a conventional interpolant for
  each of the input formulas, corresponding to the induction suggested with
  \cite[Lemma~2]{craig:uses}.

  It seems that most other implementations of Craig interpolation are on the
  basis of propositional logic with theory extensions and specialized for
  applications in verification \cite{benedikt:2017}. Craig interpolation for
  first-order logic is supported by \name{Princess}
  \cite{ruemmer:ipol:jar:2011,ruemmer:ipol:beyond:2011} and by extensions of
  \name{Vampire} \cite{vampire:interpol:2010,vampire:interpol:2012}.
  The incompleteness indicated in
  footnote~\ref{foot-craig-fol} applies to these
   \name{Vampire} extensions and was observed by their authors.
  It also appears that the \name{Vampire} extensions do not
  preserve the polarity constraints of Craig-Lyndon
  interpolants \cite{benedikt:2017}.

\section{Further Features of \PIE}  
\label{sec-further}
  
In this section we briefly describe further features of \PIE that were not
  illustrated by the examples in the previous sections.  First we consider the
  formula macro system. It utilizes Prolog variables to mimic further features
  of the processing of $\lambda$-expressions by automatically binding a Prolog
  variable that is free after computing the user-specified part of the
  expansion to a freshly generated symbol.  With a macro declaration,
  properties of its lexical environment, in particular configuration settings
  that affect the expansion, are recorded. Macros with parameters are
  processed by pattern matching to choose the effective declaration for
  expansion, allowing structural recursion in macro declarations.

    A Craig interpolant for formulas $F$ and $G$ is extracted in \PIE from a
  Prolog term that represents a closed clausal tableau, a proof of the
  validity of $F \imp G$. \PIE supports the visualization of such tableaux as
  graph, rendered by the \name{Graphviz} tool. Here is an example:
  \par\smallskip

\noindent Input: $\forall \mathit{x} \, \mathsf{p}(\mathit{x}) \land  \forall \mathit{x} \, (\mathsf{p}(\mathit{x}) \imp  \mathsf{q}(\mathit{x})) \imp  \mathsf{q}(\mathsf{c}).$\\
\noindent Result of interpolation:
\[\begin{array}{lllll}
\forall \mathit{x} \, \mathsf{q}(\mathit{x}).
\end{array}
\]
The respective directive for this interpolation task
  in the source is:\par\medskip

\noindent  
\begin{BVerbatim}[fontsize=\small]  
:- ppl_printtime(ppl_ipol((all(x, p(x)), all(x, (p(x) -> q(x))) -> q(c)),
 			  [ip_dotgraph=printstyle('/tmp/tmp01.png'),
 			   ip_simp_sides=false])).
\end{BVerbatim}
\par\medskip

  \begin{wrapfigure}[20]{R}{10.5em}
  \vspace{-0.8cm}
  \includegraphics[width=10.5em]{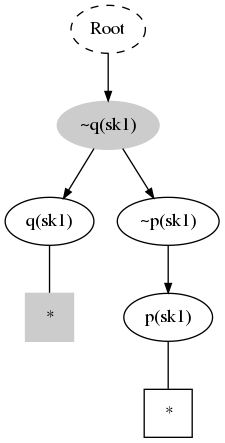}
  \caption{A clausal tableau.}
  \label{fig-tabx}
  \end{wrapfigure}
  
  \noindent The \code{ip\_dotgraph} option effects that an image representing
  the tableau is generated. The \code{ip\_simp\_sides} option suppresses
  preprocessing of the interpolation input, which, in the example, would in
  essence be already sufficient to compute the interpolant, yielding
  a trivial tableau.
  The generated image
  can then be included into the \PIE document with standard \LaTeX\ means,
  here, for example as Fig.~\ref{fig-tabx}.
  Siblings in the tableau represent a ground
  clause used in the proof.  As the tableau is used for interpolant
  extraction, decoration indicates whether the clause stems from the left or the
  right side of the input formula. The decoration
  of the closing marks indicate the side of the connection
  partner. The Skolem constant $\f{sk1}$ is converted to a quantified variable
  in a postprocessing operation. For a description of the interpolant
  extraction procedure, see \cite{cw-ipol}.

  Aside of the shown representation of quantified first-order formulas by
  Prolog ground terms, the system also supports a representation of clausal
  formulas as list of lists of terms (logic literals), with variables
  represented by Prolog variables. The system functionality can be accessed by
  Prolog predicates, also without using the document processing facilities.
  
  Practically successful reasoners usually apply in some way conversions of
  low complexity as far as possible: as preprocessing on inputs, potentially
  during reasoning, which has been termed \name{inprocessing}, and to improve
  the syntactic shape of output formulas as discussed in Sect.~\ref{sec-circ}.
  Abstracting from these situations, we subsume these conversions under
  \name{preprocessing operations}. Also the low complexity might be taken more
  or less literally and, for example, be achieved simply by trying an
  operation within a threshold limit of resources.  \PIE includes a number of
  preprocessing operations including normal form conversions, also in variants
  that produce structure preserving normalizations, various simplifications of
  clausal formulas, and an implementation of McCune's un-Skolemization
  algorithm \cite{mccune:unskolemizing}.  While some of these preserve
  equivalence, others preserve equivalence just with respect to a set of
  predicates, for example, purity simplification with respect to predicates
  that are not deleted or structure preserving clausification with respect to
  predicates that are not added.  This can be understood as preserving the
  second-order equivalence \[\exists q_1 \ldots \exists q_n\, F\; \equiv\;
  \exists q_1 \ldots \exists q_n\, G,\] where $F$ and $G$ are inputs and
  outputs of the conversion and $q_1, \ldots, q_n$ are those predicates that
  are permitted to occur in $F$ or $G$ whose semantics needs \emph{not} to be
  preserved.  If $q_1,\ldots, q_n$ includes all permitted predicates, the
  above equivalence expresses equi-satisfiability.  Some of the
  simplifications implemented in \PIE allow to specify explicitly a set of
  predicates whose semantics is to be preserved, which makes them applicable
  for Craig interpolation and second-order quantifier elimination.
  
  In addition to the implementation of the \name{DLS} algorithm, \PIE includes
  further experimental implementations of variants of second-order quantifier
  elimination. In particular, a variant of the method shown in
  \cite{lin-forget} for elimination with respect to ground atoms, which
  always succeeds on the basis of first-order logic.  A second-order
  quantifier is there, so-to-speak, just upon a particular ground
  instance of a predicate.  The \name{Boolean solution problem} or
  \name{Boolean unification with predicates} is a computational task related
  to second-order quantifier elimination
  \cite{schroeder:all,rudeanu:74,cw-boolean}.  So far, \PIE includes
  experimental implementations for special cases: Quantifier-free formulas with
  a technique from \cite{eberhard} and a version for finding solutions with
  respect to ground atoms, in analogy to the elimination of ground atoms.

\section{Conclusion}
  \label{sec-conclusion}
  
  \PIE tries to supplement what is needed to use automated first-order proving
  techniques for developing and analyzing formalizations.  Its main focus is
  not on proofs but on \emph{formulas}, as constituents of complex
  formalizations that are composed and structured through macros, and as
  computed outputs of second-order quantifier elimination, Craig interpolation
  and formula conversions that preserve semantics with respect to given
  predicates. All of these operations utilize some natural relationships
  between first- and second-order logic.

  The system mediates between high-level logical presentation and detailed
  configuration of reasoning systems: Working practically with first-order
  provers typically involves experimenting with a large and developing set of
  related proving problems, for example with alternate axiomatizations or
  different candidate theorems, and is thus often accompanied with some
  meta-level technique to compose and relate the actual proof tasks submitted
  to first-order reasoners.  With the macro system, the supported
  document-oriented workflow, \LaTeX\ pretty-printing, and integration into
  the Prolog environment, \PIE offers to organize this in a systematic way
  through mechanisms that remain in the spirit of first-order logic, which in
  mathematics is actually often used with schemas.

  Aside of the current suitability for non-trivial applications, \PIE shows up
  a number of challenging and interesting open issues for research, for
  example improving practical realizations of second-order quantifier
  elimination, strengthenings of Craig interpolation that ensure
  application-relevant properties such as range restriction, and conversion of
  computed formulas that are basically just semantically characterized to
  comprehensible presentations.  Progress in these issues can be directly
  experienced and verified with the system.

\bibliographystyle{splncs04}
\bibliography{bib_pie_01}
\end{document}